\documentstyle[preprint,aps]{revtex}
\begin{document}

\title{Can dileptons reveal the in-medium properties of vector mesons?}
\bigskip
\author{G. Q. Li and C. M. Ko}
\address{Cyclotron Institute and Physics Department,\\  Texas A\&M University,
College Station, Texas 77843, USA}

\maketitle

\begin{abstract}
Dilepton production from both pion-pion and kaon-antikaon annihilation
in heavy-ion collisions is studied using the relativistic transport
model. The formation of a rho meson from pion-pion annihilation and a phi
meson from kaon-antikaon annihilation, their propagation in the medium,
and their decay into dileptons
are explicitly treated.
Including the medium modifications of the masses and widths of vector mesons
as predicted by the QCD sum-rule calculations, we
study their effects on the dilepton invariant mass spectra from heavy-ion
collisions at SIS/GSI
energies.
\end{abstract}


\medskip
\section{Introduction}

Hadron properties are believed to be modified in the nuclear medium due
to the partial restoration of chiral symmetry. A well-known example is
the reduced nucleon mass in nuclear matter, as shown in
the Walecka model \cite{QHD}, in the Dirac-Brueckner calculation
\cite{SHA83,MAL87,MAC90,LI92}, and more recently in the QCD sum-rule
calculations \cite{COH91,COH92}. Based on the scaling
properties of QCD, Brown
and Rho \cite{BRO91} have argued that the masses of non-strange vector mesons
in nuclear matter
should also be reduced at about the same rate
as the nucleon mass. This intriguing
suggestion has led to many theoretical studies
of hadron in-medium properties based on the QCD
sum rules \cite{HAT92A,HAT92B,ASA93} and the effective field theory
\cite{PIE93,HAT94,SONG94}.

Phenomenologically, the enhanced cross section in kaon-nucleus scattering and
the suppression of the electromagnetic longitudinal response in
electron-nucleus
scattering have been considered as evidence for the decrease of vector
meson masses in the nuclear medium \cite{BRO88,BRO89}.
However, the density involved in these
reactions is small, i.e.,
below or at most equal to the normal nuclear matter density.
On the other hand,
heavy-ion collisions at intermediate
energies provide a unique way to form in the laboratory
a piece of {\it dense matter} with
density up to 2-3$\rho _0$ ($\rho _0\approx$ ~0.16 fm$^{-3}$) and
thus make it possible to study experimentally the properties of hadrons
in a dense medium.
Unfortunately, the dense matter formed in heavy-ion collisions
exists only as an intermediate stage for a very short period. Direct
experimental studies of the dense medium is thus impossible. A
theoretical transport model has to be used in order to deduce the
properties of the intermediate stage from the known initial conditions and
the measured final observables. Among transport models,
the relativistic transport model based on the Walecka model
offers a consistent description of changing hadron masses in
the dense matter formed in intermediate-energy
heavy-ion collisions \cite{KO87,KO88,LI89,MOS88}.
In this model,
the nucleon mass in a dense medium is reduced by the attractive
scalar field, and the
energy is thus stored in the scalar field energy.
As the density decreases, the scalar field energy diminishes
and is converted back to the nucleon mass, which
eventually returns to its free mass at freeze-out.
In addition to the propagation of nucleons (and deltas)
in the mean-field potential,
the relativistic transport model also incorporates elastic and
inelastic two-body collisions among nucleons, deltas, and pions.

In order to determine unambiguously the properties of hadrons in
a dense matter, one needs also to find suitable observables
that carry the information of the intermediate stage and are not
significantly affected by final-state interactions. The electromagnetic
signals, especially dileptons,
due to their weak interaction with hadrons, have been
considered as good observables to probe the properties of the
intermediate stage of heavy-ion collisions.
There have already
been studies of dilepton production in heavy-ion collisions
using both fire-ball models \cite{GAL87,GAL88,XIA88,KO89,KOR90A,KOR90B}
and transport models \cite{XIO90A,XIO90B,WOL91,WOL93}. In these
studies, it has been shown that dileptons are indeed mostly produced
from the hot and dense stage of heavy-ion collisions.

The possibility of
studying in-medium pion dynamics through dilepton production
from pion-pion annihilation was first pointed out in Ref. \cite{GAL87}
in a schematic model and later confirmed in Ref. \cite{XIA88} using
the delta-hole model.
Because of the softening of the pion dispersion relation,
an enhanced yield of dileptons
with invariant masses around twice the pion mass was predicted.
However, more refined investigations indicate that
this enhancement is largely cancelled by vertex corrections
\cite{KOR90A,KOR90B}.
Furthermore,
for dileptons with invariant masses around twice the pion mass, the background
from the eta and delta Dalitz decays as well as the neutron-proton
bremsstrahlung are more significant than that from the pion-pion
annihilation \cite{WOL93}. It is thus very difficult, if  not impossible, to
study the in-medium pion dispersion relation from measuring the dilepton
invariant mass spectrum.

Since the electromagnetic form factor of a pion is dominated by the rho meson
according to the vector dominance model,
the dilepton spectrum may also reveal information about the property
of rho meson in dense matter. For large dilepton invariant masses,
the background from Dalitz decays and bremsstrahlung is small, it should
therefore be easy to identify the rho meson. Calculations using the free
rho-meson mass and width in the  pion electromagnetic form factor
indeed reveal a broad but visible peak in the
dilepton spectrum at invariant masses around the free rho-meson mass ($\approx$
770 MeV) \cite{XIO90B,WOL91}.
In Ref. \cite{NOR93}, the medium modification of the pion
electromagnetic form factor due to the delta-hole polarization
has been studied.  As a result of the softened pion dispersion relation
in a medium,
the pion electromagnetic form factor is broadened. Its effect on the
dilepton spectra has been investigated in Ref. \cite{WOL93}, and it
has been found
that the peak at the rho-meson mass almost disappears because of the increase
of the rho-meson width in the medium.
However, the calculation of Ref. \cite{NOR93} neglects the
medium effects on the vacuum,
which, when included as in the QCD sum-rule \cite{HAT92A,HAT92B}
and effective field theory calculations \cite{HAT94}, lead to the
reduction of the rho-meson mass in dense matter.
The rho-meson peak in
the dilepton spectrum is thus expected to shift to lower invariant masses.

Another vector meson whose properties in a
dense medium have been under extensive
discussion is the phi meson \cite{HAT92A,HAT92B,SHU91,SHU92,KO92,KO94,ASA94}.
The QCD sum-rule calculation shows that
the phi-meson mass also decreases with increasing density, but at a smaller
rate than that of the rho meson \cite{HAT92A,HAT92B}.
The width of the phi meson also
changes with density, mainly due to the medium modification of kaon and
antikaon masses, as well as the phi-meson mass, as will be discussed later.
Since the kaon-antikaon annihilation to a dilepton mainly
proceeds through the phi meson,
dilepton production in heavy-ion collisions thus allows one
to learn about the properties of the phi meson in the
dense medium.

The main purpose of this paper is to study
dilepton production from both pion-pion and  kaon-antikaon annihilation
in heavy-ion collisions to see whether one can learn about
the in-medium properties
of the rho and phi mesons from the dilepton spectrum.
For the purpose of illustration we shall use in our calculation
the in-medium masses
of the rho and phi mesons as predicted by the QCD sum-rule calculations
of Refs. \cite{HAT92A,HAT92B}. It should be mentioned that due to
ambiguities in the four-quark condensate and in the strangeness
content of a nucleon, the present QCD sum-rule predictions for
the in-medium masses of the rho and phi mesons are not without
uncertainty.

In all previous calculations of dilepton production from the pion-pion
(kaon-antikaon) annihilation
\cite{GAL87,GAL88,XIA88,KO89,KOR90A,KOR90B,XIO90A,XIO90B,WOL91,WOL93},
the effects of the intermediate
rho (phi) meson are included through the pion (kaon)
electromagnetic form factor using the vector dominance model. In
this form factor
approach, the time delay between the formation of the rho (phi) meson
and its decay into a dilepton is neglected.
Physically, dilepton production from the pion-pion (kaon-antikaon)
annihilation proceeds in three steps. First, a rho (phi) meson is formed
from the pion-pion (kaon-antikaon) annihilation, which then propagates in
the hadronic medium, and finally decays into a dilepton with an appropriate
branching ratio. In this
dynamical approach, the formation and decay of vector mesons  are
treated explicitly. If
there are no medium effects, the two approaches give essentially the same
results. Since we are interested in possible medium effects
on the vector mesons in the dilepton spectra, the dynamical
approach is more realistic. Neglecting the time delay between
formation and decay  of vector mesons
(i.e., neglecting dilepton formation time) in the
form factor approach would overestimate the medium effects.

This paper is organized as follows. In section 2, we briefly  review the
relativistic transport model of Ref. \cite{KO87,KO88,LI89}. The necessary
formalism for calculating dilepton production from both pion-pion and
kaon-antikaon annihilation is discussed in section 3.
The results are presented in section 4. Finally, a  short summary is
given in section 5.

\section{Relativistic transport model}

Our study is based on the relativistic transport model developed in Refs.
\cite{KO87,KO88,LI89}. At energies considered in this work, the colliding
system consists mainly of nucleons, deltas and pions. Pions are
produced from decays of deltas that are excited in
nucleon-nucleon inelastic interactions.
While pions are treated
as free particles, nucleons and deltas are propagated in their
mean-field potentials according to the following equations of motion
\begin{eqnarray}
{d{\bf x}\over dt}~=~{\bf p}^*/E^*,~~{d{\bf p}\over dt}~=~-\nabla _x
\big[E^*+(g_\omega /m_\omega )^2\rho _B\big],
\end{eqnarray}
with $E^*=(m^{*2}+{\bf p}^{*2})^{1/2}$. The effective mass and kinetic
momentum of a baryon (nucleon or delta) are given, respectively,
by $m^*=m+\Sigma _S$,
and ${\bf p}^*=~{\bf p}-{\bf \Sigma}_V$, where the scalar $\Sigma _S$
and the vector ${\bf \Sigma}_V$ self-energies
are determined from the non-linear $\sigma$-$\omega$ model.
We use the so-called soft equation-of-state, as determined in Ref.
\cite{LI94A}, in the present calculation.

Elastic ($NN\rightarrow NN, N\Delta\rightarrow N\Delta ,
\Delta\Delta\rightarrow \Delta\Delta$) and inelastic ($NN\leftrightarrow
N\Delta , \Delta\leftrightarrow N\pi$) reactions among nucleons,
deltas and pions are also included. The standard Cugnon
parametrizations  \cite{CUG81,BER88}
and the proper detailed-balance prescription \cite{DAN91} are used for
describing these reactions. In Fig. 1,
we show the time evolution of the average central density
(dotted curve) and the pion multiplicity for a Ni+Ni collisions at an incident
energy of 2.0 GeV/nucleon and an impact parameter of 0 fm.
The final pion multiplicity is about 39, which gives a pion-to-nucleon ratio
of about 0.33, and is in good agreement with the experimental data
\cite{HAR87}.
We also see that pions materialize at a later stage of the collision
when the system has already started to expand.

For heavy-ion collisions at incident energies considered here,
kaon and antikaon production are treated perturbatively
as their production probabilities are very small. We refer to Refs.
\cite{FAN94,LI94B} for details on the elementary production
cross section and the treatment of final-state interactions. In Refs.
\cite{FAN94,LI94B}, we have found that an attractive scalar potential
derived from the chiral Lagrangian \cite{KA86,KA87,BRO87,RHO93} is
needed for both the kaon and the
antikaon in order to explain quantitatively the
experimental data from SIS/GSI \cite{GSI1,GSI2}.
In the mean-field approximation to the chiral Lagrangian,
the effective masses of the kaon and the antikaon in a nuclear medium are
\begin{eqnarray}
m_{K,{\bar K}^*}=
m_K\Bigg[1-{\Sigma _{KN}\over f_K^2m_K^2}\rho _S+\Big({3\over 8}
{\rho _B\over f_K^2m_K}\Big)^2\Bigg]^{1/2}\pm {3\over 8}{\rho _B\over f_K^2},
\end{eqnarray}
where the plus and the minus sign correspond to the kaon and the antikaon,
respectively.
With a kaon decay constant $f_K\approx 93$ MeV and a $KN$ sigma term
$\Sigma _{KN}\approx 350 $ MeV,
we find that the kaon mass increases
with density while the antikaon mass decreases with density,
as shown in Fig. 2. In Fig. 1,
we also show the time evolution of the kaon
and antikaon multiplicities. Kaons and antikaons are
produced in the early stage of the collision when baryon-baryon
collisions are most energetic. Their abundance reaches a maximum when
the system is most compressed. The final kaon multiplicity is about 0.33,
and that of the antikaon is about 0.009.
We note that if the attractive scalar
potential for the kaon and the antikaon is neglected, the kaon yield is reduced
by about 40\% to 0.22 (dashed curve),
while the antikaon yield is reduced by about a
factor of 3 to 0.003 (dashed curve).

\section{dilepton production: formalism}

\subsection{Pion-pion annihilation}

According to  the vector dominance model, dilepton production from pion-pion
annihilation proceeds through a rho meson which then converts to a virtual
photon and decays into a dilepton.  This is shown by the
diagram in Fig. 3. Including the medium
modification of the rho-meson mass and width which will be specified later,
the invariant matrix element for this process is
\begin{eqnarray}
{\cal M} ={e^2 m_\rho ^{*2}\over q^2\Big[q^2-\big(m_\rho ^*-{i\over 2}
\Gamma ^*_\rho\big)^2\Big]}
(\not\mkern-4mu q_1-\not\mkern-4mu q_2){\bar u}(k_1,s_1)v(k_2,s_2),
\end{eqnarray}
where we have used the relation,
$g_{\rho\gamma}={em_\rho ^{*2}\over g_{\rho\pi\pi}}$, between
the $\rho\pi\pi$ coupling constant $g_{\rho\pi\pi}$
and the $\rho\gamma $ coupling constant $g_{\rho\gamma}$
\cite{BHA}.

{}From Eq. (3) we have
\begin{eqnarray}
{\bar {\cal M}^2}={2e^4m_\rho ^{*4}\over m_e^2q^4\Big[\big(q^2-m_\rho
^{*2}\big)^2+\big(m_\rho^*\Gamma _\rho^*\big)^2\Big]}
\Big[-{1\over 4}q^4+m^2_\pi q^2+4\big({\bf k\cdot q}\big)^2\Big].
\end{eqnarray}
The dilepton production cross section from pion-pion annihilation is
then given by
\begin{eqnarray}
\sigma _{\pi^+\pi^-\rightarrow \rho^0\rightarrow e^+e^-}(M)
= {8\pi \alpha ^2k\over 3M^3}{m_\rho ^{*4}\over \big(M^2-m_\rho^{*2}\big)^2
+\big(m_\rho ^*\Gamma _\rho ^*\big)^2},
\end{eqnarray}
where $M$ is the invariant mass of the $\pi ^+\pi ^-$ pair, and $\alpha$
is the fine structure constant. The three momentum $k$ is given by
\begin{eqnarray}
k={1\over 2}\big(M^2-4m_\pi ^2\big)^{1/2}.
\end{eqnarray}

Theoretical studies based on the QCD sum rules \cite{HAT92A,HAT92B} and
the
effective field theory \cite{HAT94} indicate that the rho-meson mass decreases
with increasing nuclear density. In this work, we use the
results from the QCD sum-rule calculation \cite{HAT92A,HAT92B}
which can be parametrized by
\begin{eqnarray}
{m^*_\rho\over m_\rho}\approx 1-0.18({\rho/ \rho _0})
\approx {1\over 1+0.18({\rho/ \rho _0})}.
\end{eqnarray}
The first parametrization has been given in Refs. \cite{HAT92A,HAT92B}.
We rewrite it as the second form to extend to higher densities.
The in-medium rho-meson mass is seen to
decrease with increasing density, as shown in
Fig. 2. At twice the normal nuclear matter density,
it decreases by about 25\%.
We note that in the QCD sum-rule calculations, the density dependence of
the rho-meson mass is largely determined by the density dependence of
the four-quark condensate which has been, however,
only approximately calculated.

The width of the rho meson in a medium is given by
\begin{eqnarray}
\Gamma ^*_\rho ={g^2_{\rho \pi \pi}\over 4 \pi} {1\over 12m^{*2}_\rho }
(m^{*2}_\rho -4m^2_\pi )^{3/2},
\end{eqnarray}
where the $\rho\pi\pi$
coupling constant $g_\rho\pi\pi$ is determined from the decay
width of the rho meson in free space. Using $m_\rho$=768.1 MeV, $m_\pi$=139.6
MeV and $\Gamma _\rho$=151.5 MeV \cite{DATA}, we have  $g^2_{\rho\pi\pi}/4\pi
\approx 2.9$. Because of the reduction of the rho-meson mass in the medium,
the width of the rho meson decreases with increasing
density, as shown in Fig. 4.
In Eq. (8), the effect on the rho-meson width due to the medium modification of
the pion dynamics has not been included,
as it has been shown to be less important than the
effect from the decrease of the rho-meson mass in the medium \cite{ASA93}.
Also, the possible medium modification of the
$\rho\pi\pi$ coupling constant has been
neglected.

In order to treat explicitly the formation, propagation and
decay of the rho meson,
we express the dilepton production cross section, Eq. (5), as the
product of the rho-meson formation cross section and the branching-ratio
for its decay into a dilepton, i.e.,
\begin{eqnarray}
\sigma _{\pi^+\pi^-\rightarrow \rho^0\rightarrow e^+e^-}(M)
=\sigma _{\pi^+\pi^-\rightarrow \rho^0}(M){\Gamma _{\rho\rightarrow e^+e^-}
(M).
\over \Gamma _\rho (M)},
\end{eqnarray}
where $\Gamma _{\rho\rightarrow e^+e^-} (M)$ is the rho-meson
decay width into a dilepton.
Using the following expressions  \cite{BHA}
\begin{eqnarray}
\Gamma _{\rho\rightarrow
e^+e^-}(M)={1\over 3}\alpha ^2M\Big({4\pi\over g^2_{\rho\pi\pi}}
\Big),
\end{eqnarray}
\begin{eqnarray}
\Gamma _\rho (M)={2\over 3}{g^2_{\rho \pi \pi}\over 4 \pi} {k^3\over M^2},
\end{eqnarray}
and comparing Eq. (9) with Eq. (5), we have
\begin{eqnarray}
\sigma _{\pi^+\pi^-\rightarrow \rho ^0}(M)={12\pi\over k_\rho ^{*2}}
{(m^*_\rho\Gamma ^*_\rho)^2
\over (M^2-m^{*2}_\rho )^2+(m^*_\rho\Gamma ^*_\rho )^2}
\big({k\over k_\rho ^*}\big)^4
\big({m_\rho ^*\over M}\big)^6,
\end{eqnarray}
where $k_\rho ^*$ is given by Eq. (6) with $M=m^*_\rho$.
This is the rho-meson formation cross section used in our numerical
calculation.

The propagation of the rho meson is treated as that of a free particle,
i.e., we neglect the effect of the mean-field potential during its propagation.
We include, however,  approximately
the change of the rho-meson mass as it propagates
from one place (with density $\rho _1$) to another (with density $\rho _2$):
\begin{eqnarray}
M_2=M_1+m_\rho \Big[{0.18\big(\rho _1/\rho _0\big)\over
1+0.18\big(\rho _1/\rho _0\big)}-{0.18\big(\rho _2/\rho _0\big)\over
1+0.18\big(\rho _2/\rho _0\big)}\Big].
\end{eqnarray}
This is similar to the change of the nucleon effective mass as it
propagates through  matter via the change of the scalar field.

Denoting, at time $t$,  the differential multiplicity
of rho mesons with mass $M$ by $P_\rho (M,t) $,
then the differential dilepton production probability is given by
\begin{eqnarray}
P_{e^+e^-}(M)=~\int _0^T P_\rho (M,t) \Gamma _{\rho\rightarrow e^+e^-} (M)dt
+P_\rho (M,T) {\Gamma _{\rho\rightarrow e^+e^-}(M)\over \Gamma _\rho (M)},
\end{eqnarray}
where $\Gamma _{\rho\rightarrow
e^+e^-} (M) $ and $\Gamma _\rho (M)$ are given by
Eqs. (10) and (11), respectively. In our numerical simulation,
$T$ in Eq. (14) is taken to be 32 fm/c. At that time,
the central density of the system is about 0.01$\rho_0$, thus any
further change of the medium effects can be neglected, and the
contribution of any remaining rho mesons to dileptons is
calculated using its branching ratio as shown by the second
term in Eq. (14).

In the form factor approach, the effect of the intermediate rho meson
is included through the pion electromagnetic form factor,
\begin{eqnarray}
|F_\pi(M)| ^2= {m_\rho ^{*4}\over (M^2-m^{*2}_\rho )^2+m^{*2}_\rho
\Gamma _\rho ^{*2}}.
\end{eqnarray}
The dilepton production cross section in this approach is
then given by
\begin{eqnarray}
\sigma _{\pi ^+\pi ^-\rightarrow e^+e^-} (M) = {8\pi \alpha ^2k\over
3M^3}|F_\pi(M)|^2.
\end{eqnarray}
This is the same as Eq. (5). Neglecting medium effects, Eq. (16)
reduces to the familiar formula used in Refs.
\cite{GAL87,GAL88,XIA88,KO89,KOR90A,KOR90B,XIO90A,XIO90B,WOL91,WOL93}.

\subsection{Kaon-antikaon annihilation}

Dilepton production from kaon-antikaon annihilation proceeds mainly
through a phi meson which couples to a virtual photon
and decays into a dilepton,
as shown by the diagram in Fig. 5.
Including the medium modifications of the kaon, antikaon
and phi meson masses and the phi width, the invariant matrix element for
this process is
\begin{eqnarray}
{\cal M} ={e^2 m_\phi ^{*2}\over 3q^2\Big[q^2-\big(m_\phi ^*-{i\over 2}
\Gamma ^*_\phi\big)^2\Big]}
(\not\mkern-4mu q_1-\not\mkern-4mu q_2){\bar u}(k_1,s_1)v(k_2,s_2),
\end{eqnarray}
where the relation,
$g_{\phi\gamma}= {em_\phi ^{*2}\over 3g_{\phi K{\bar K}}}$,
between the $\phi K{\bar K} $ coupling constant $g_\phi K {\bar K}$
and the $\phi\gamma $ coupling constant $g_\phi\gamma$ has been used
\cite{BHA}.

{}From Eq. (17), we have
\begin{eqnarray}
{\bar {\cal M}^2}={e^4m_\phi ^{*4}\over 9m_e^2q^4\Big[\big(q^2-m_\phi
^{*2}\big)^2+\big(m_\phi^*\Gamma _\phi^*\big)^2\Big]}\nonumber
\end{eqnarray}
\begin{eqnarray}
\cdot \Big[-{1\over 2}q^4+(m_K^{*2}+m_{\bar K}^{*2})q^2
+8\big({\bf k\cdot q}\big)^2
-{1\over 2}\big(m_K^{*2}-m_{\bar K}^{*2}\big)^2\Big].
\end{eqnarray}
The dilepton production cross section from the kaon-antikaon annihilation is
then given by
\begin{eqnarray}
\sigma _{K{\bar K}\rightarrow \phi\rightarrow e^+e^-}(M)
= {8\pi \alpha ^2k\over 27M^3}{m_\phi ^{*4}\over \big(M^2-m_\phi^{*2}\big)^2
+\big(m_\phi ^*\Gamma _\phi ^*\big)^2},
\end{eqnarray}
where $M$ is the invariant mass of the kaon-antikaon pair,
and the three momentum $k$ is given by
\begin{eqnarray}
k={1\over 2M}\Bigg[\Big(M^2-
(m_K^*+m^*_{\bar K})^2\Big)\Big(M^2-(m^*_K-m^*_{\bar K})^2\Big)\Bigg]
^{1/2}.
\end{eqnarray}

The QCD sum-rule calculations \cite{HAT92A,HAT92B}
indicate that the phi meson mass is also modified in the medium.
The results of Refs. \cite{HAT92A,HAT92B}, obtained
with a nucleon strangeness content of $y=2<{\bar s}s>_N/<{\bar u}u+{\bar d}d>_N
\approx 0.17$,  can be
parametrized by the simple expression
\begin{eqnarray}
{m^*_\phi\over m_\phi}\approx {1-0.0255({\rho/ \rho _0})}
\approx {1\over 1+0.0255({\rho/ \rho _0})},
\end{eqnarray}
which shows that the
phi-meson mass decreases slowly with increasing density, as shown in Fig. 2.
At twice the normal nuclear matter density, the phi-meson mass decreases
by about 5\%. We note that in the QCD sum-rule calculations,
the in-medium mass of the phi meson depends on the nucleon
strangeness content which is, however,  not well determined.

The decay width
of the phi meson into a kaon and an antikaon is  given by
\begin{eqnarray}
\Gamma ^*_\phi = {g^2_{\phi K \bar K}\over 4 \pi} {1\over 6m_\phi ^{*5}}
\Bigg[ \Big(m_\phi ^{*2}-(m^*_K+m^*_{\bar K})^2\Big)\Big(m_\phi^{*2}-(m^*_K
-m^*_{\bar K})^2\Big)\Bigg]^{3/2}.
\end{eqnarray}
As in the case of the rho meson,
the coupling constant $g_{\phi K \bar K}$ is assumed not to be modified
in the medium and is determined from the
decay width of the phi meson in free space
into a kaon and an antikaon.
Using $m_K$=495.5 MeV,
$m_\phi$=1019.4 MeV and $\Gamma _{\phi\rightarrow K\bar K}$=3.7 MeV
\cite{DATA}, we obtain
$g^2_{\phi K \bar K}/4\pi$=1.69. We note that in the present calculation
the small decay width of the phi meson into a pion and a rho meson is
neglected.
The density dependence of the phi decay width
calculated from Eq. (22) is shown in Fig. 3 as a function of density.
Although the phi meson mass is reduced in the medium,
its width
increases with density as a result of the larger reduction of the antikaon
in-medium mass than the increase of the kaon in-medium mass
and hence an increase of the available phase space. At twice the
normal nuclear matter density, the phi-meson decay width is about 30 MeV.

Similarly, to treat explicitly the formation, propagation
and decay of the phi meson,
we express Eq. (19) as the product of the phi-meson formation
cross section and the branching ratio for its decay into the
dilepton, i.e.,
\begin{eqnarray}
\sigma _{K{\bar K}\rightarrow \phi\rightarrow e^+e^-}(M)
=\sigma _{K{\bar K}\rightarrow \phi}(M){\Gamma _{\phi\rightarrow e^+e^-} (M)
\over \Gamma _\phi (M)}.
\end{eqnarray}
where $\Gamma _{\phi\rightarrow e^+e^-} (M)$ is the phi-meson decay width into
the dilepton.
Using the following expressions
\begin{eqnarray}
\Gamma _{\phi\rightarrow
e^+e^-}(M)={1\over 27}\alpha ^2M\Big({4\pi\over g^2_{\phi K{\bar K}}}
\Big),
\end{eqnarray}
\begin{eqnarray}
\Gamma _\phi (M)={4\over 3}{g^2_{\phi K{\bar K}}\over 4 \pi} {k^3\over M^2},
\end{eqnarray}
and comparing Eq. (23) with (19), we get
\begin{eqnarray}
\sigma _{K{\bar K}\rightarrow \phi}(M)={6\pi\over k_\phi ^{*2}}
{(m^*_\phi\Gamma ^*_\phi)^2
\over (M^2-m^{*2}_\phi )^2+(m^*_\phi\Gamma ^*_\phi )^2}
\Big({k\over k_\phi ^*}\Big)^4
\Big({m_\phi^*\over M}\Big)^6,
\end{eqnarray}
where $k_\phi ^*$ is given by Eq. (20) with $M=m_\phi ^*$.

The propagation of the phi meson is also treated as that of a free
particle. The change of its mass in the medium
as it propagates from one place  to another is approximately described by
\begin{eqnarray}
M_2=M_1+m_\rho \Big[{0.0255\big(\rho _1/\rho _0\big)\over
1+0.0255\big(\rho _1/\rho _0\big)}-{0.0255\big(\rho _2/\rho _0\big)\over
1+0.0255\big(\rho _2/\rho _0\big)}\Big],
\end{eqnarray}
as in the case of the rho meson.

The differential dilepton production probability is then given by
an equation similar to Eq. (14), with $\Gamma _\rho (M),
{}~\Gamma _{\rho\rightarrow e^+e^-} (M)$, and $P_\rho
(M,t)$ replaced, respectively,
by  $\Gamma _\phi (M),
{}~\Gamma _{\phi\rightarrow e^+e^-} (M)$, and $P_\phi(M,t)$, where
$P_\phi (M,t)$ is the
differential multiplicity at time $t$ of phi mesons with mass M.

In the form factor approach, the effect of the intermediate phi meson
is included through the kaon electromagnetic form factor \cite{GAL88,REN77},
\begin{eqnarray}
|F_K(M)| ^2\approx
{1\over 9}{m_\phi ^{*4}\over (M^2-m^{*2}_\phi )^2+m^{*2}_\phi
\Gamma _\phi ^{*2}}.
\end{eqnarray}
The cross section for the production of dileptons with an invariant mass $M$
from the kaon-antikaon annihilation
is then given by
\begin{eqnarray}
\sigma ^{e^+e^-}_{K{\bar K}} (M) = {8\pi \alpha ^2k\over
3M^3}|F_K(M)|^2.
\end{eqnarray}
This is the same as Eq. (19) and reduces to the formula used in
Ref. \cite{GAL88}
when the medium effects are neglected.

\section{dilepton production: results and discussions}

In obtaining the dilepton invariant mass spectra,
we have used a bin size of 10 MeV for dileptons from
pion-pion annihilation
and a bin size of 4 MeV for those from kaon-antikaon annihilation.
We have used 100 test particles for each run of the simulation,
and repeated the simulation 40 and 10 times for
calculating the pion-pion and the kaon-antikaon contribution, respectively.
For pion-pion annihilation, we have thus more than 1000 events
in each low invariant mass bin, and about 100 events in each
high invariant mass  bin. For kaon-antikaon
annihilation, we have a few thousand events in each low
invariant mass bin, and a few hundred in
each high invariant mass  bin.  We note that
in our perturbative
treatment of kaon and antikaon production, we allow a quasi-kaon
or quasi-antikaon to be produced whenever the available energy of a
baryon-baryon collision is above the kaon or the antikaon production
threshold. This quasi-kaon or quasi-antikaon carries a probability which
is given by the ratio of its production cross section to the baryon-baryon
total cross section. We find that  at the time when the system is most
compressed, there are about 150 quasi-kaons and 30 quasi-antikaons, while the
(physical) pion number is about 20.
In this way, we obtain good statistics for both  pion-pion and
kaon-antikaon annihilation.

We first carry out a calculation in which the medium effects
on kaons, antikaons, rho and phi mesons are neglected, i.e., we use
free masses and widths in Eqs.  (12) and (26). Also, in calculating
the kaon and antikaon production probabilities
the attractive scalar potential is
neglected. The total dilepton spectrum from both pion-pion  and
kaon-antikaon annihilation is
shown in Fig. 6.
Since in our numerical simulation a bin size of 10 MeV
is used for dileptons from  pion-pion annihilation, it has been converted
to a new bin size of 4 MeV by interpolations in order to add it to the
kaon-antikaon contribution.

For pion-pion annihilation
we observe a broad but visible peak around the rho-meson
mass ($\approx$ 770 MeV),
which is in agreement with earlier results
obtained in non-relativistic transport models \cite{XIO90B,WOL91}.
Since most
pion-pion collisions are of low invariant masses,
a bump is thus seen in this region. As shown in Refs.
\cite{WOL91,WOL93}, the background
from the eta and delta
Dalitz decays  and the neutron-proton bremsstrahlung is much smaller
than the pion-pion contribution around the rho-mass region.
Therefore, if the rho meson mass is not modified in the
medium, one should observe a broad peak around 770 MeV in the
dilepton spectrum.

For kaon-antikaon annihilation we
observe a very sharp peak around the phi-meson
mass ($\approx $1.02 GeV). This peak
is about a factor of 4 above the background from
pion-pion annihilation. At these invariant masses, the background
from other sources, e.g., Dalitz decays and bremsstrahlung,
are again small or at most comparable to that from  pion-pion annihilation
\cite{WOL91,WOL93}. Thus, if the properties of the kaon, the antikaon and
the phi meson
are not modified in the medium, one should see a peak in the
dilepton spectrum around 1020 MeV. Of course,
a detector of excellent resolution is needed to see
this peak due to its very narrow width ($\approx 4 $MeV).

Our main point, however, is to see
whether we can learn from  dilepton spectra
about  medium modifications of hadron properties.
We show in Fig. 7 the dilepton spectrum
obtained by including in the calculation
all medium effects discussed above.
We find that the maximum in the
pion-pion contribution shifts from the free rho-meson mass to around
550 MeV as a result of the decrease of the rho-meson mass in dense
matter. The height of this peak  is about four times the
height of the peak at the free rho-meson mass when the  medium effects
are not included.
This very broad peak results
from the overlap
of a number of peaks with different positions and widths (depending on the
density where the pion-pion annihilation occurs). The small shoulder
at low invariant masses is again due to the fact that  most pion-pion
collisions
are of such invariant masses.
According to Ref. \cite{WOL93},
the background from
Dalitz decays and bremsstrahlung around 550 MeV is, at
most, of the same magnitude as that of the free rho-meson peak obtained without
the medium effects.
Since the rho-meson
peak at 550 MeV  after including the medium effects
is enhanced by about a factor of four as compared to the
original peak without the medium effects, we expect that if
the rho-meson mass
is indeed reduced in the medium as predicted by the QCD sum-rule
calculations, a broad peak in the dilepton
invariant mass spectrum around 550 MeV will be observed.
Although to detect such a broad peak
is difficult experimentally, the study of the in-medium properties of
the rho meson from the dilepton spectra is still possible if the background
from Dalitz decays and bremsstrahlung can be removed.

The dilepton spectrum from  kaon-antikaon annihilation is also
modified when the medium effects are included. The  peak now
shifts to a  slightly lower invariant mass
and is about a factor of five broader than the original
peak at the free phi-meson mass when no medium effects are included.
The broadening of the phi peak is mainly due
to the increase of its width in the medium and the overlap
of a number of peaks with different widths and positions.
More interestingly,  a raised shoulder appears around 950 MeV as a result
of the
decrease of the phi-meson mass at high densities.
This shoulder is about a factor of two above the
background from pion-pion annihilation. The background from other
sources of dileptons at these invariant masses,
such as Dalitz decays and bremsstrahlung,
is small. It is thus feasible
to study the in-medium properties of the phi meson
by measuring  dilepton spectra in heavy-ion collisions.
Finally, because of the decrease
of the antikaon mass in the medium, we have dileptons with invariant
masses as low as 800 MeV from the kaon-antikaon annihilation.

The change of the dilepton spectrum due to the
medium effects can be more clearly seen
by examining the pion-pion and the kaon-antikaon contribution separately.
In Fig. 8, we compare the dilepton spectra
from pion-pion annihilation with and without medium effects on
the rho meson. The shapes of the spectra are clearly different. The
peak shifts to lower invariant masses and becomes broader.
In Fig. 9, a similar comparison is shown for the dilepton
spectra from kaon-kaon annihilation. The appearance of a shoulder around
950 MeV and the broadening of the peak around the free phi-meson mass
are clearly observed.

To test the reliability of the form factor approach
that has been used often in previous calculations, we have also carried out two
calculations using Eqs. (16) and (29).
The results are shown in Fig. 10 and Fig. 11 for the cases without medium
effects and with medium effects, respectively.
Comparing Fig. 10 with Fig. 6, we see that if the medium effects are neglected,
the form factor approach gives essentially the same results as those
obtained from the
dynamical approach for both pion-pion and kaon-antikaon annihilation.
When the medium effects are included,
the two methods of
calculations lead to similar predictions for the dilepton spectra
from pion-pion annihilation, i.e., the shift of the peak from
the free rho-meson mass
to a lower invariant mass and the broadening of the peak when the medium
effects are included. This similarity can be easily understood, as the
decay width of a rho meson is quite large (see Fig. 4 where the decay width
corresponding to the peak mass $m_\rho ^*$ is shown) and the rho meson
propagates thus for only a very short distance before it decays. The neglect
of the formation time of dileptons in the form factor approach is then not
a bad approximation. However, there do exist some quantitative
differences between the results of the two calculations, namely, the peak
shifts to a slightly larger invariant mass and its width is almost doubled
in the case of the dynamical approach as compared to the results from
the form factor approach.

The situation is quite different for kaon-antikaon annihilation.
In the form factor
approach, we see in Fig. 11 a peak around 950 MeV which is
well above the background from pion-pion annihilation, whereas
in the dynamical approach
only a raised shoulder is seen to be just above this background.
On the other hand, in the form factor approach we also have a small peak around
the free phi-meson mass that comes from kaon-antikaon annihilation
at the later stage of heavy-ion collisions when the density is low.
In the dynamical approach, there appears a
strong peak just slightly
below the free phi-meson mass. The apparent
difference between the results of the two calculations goes back to
the fact that the phi-meson width is small (see Fig. 4).
A phi meson, when formed from the kaon-antikaon annihilation at high densities,
does not decay right away as assumed in the
form factor approach, but rather propagates in the medium to lower
densities as the system expands. Therefore, for those intermediate
vector mesons with long life times, the explicit treatment
of their formation, propagation and decay is necessary if we want
to study their in-medium properties by measuring the dilepton invariant mass
spectra.

\section{summary}

In summary, we have studied dilepton production
from both pion-pion and kaon-antikaon annihilation
in central heavy-ion collisions at SIS/GSI energies.
The collision dynamics have been described by the generalized relativistic
transport model which includes the medium effects on kaons and
antikaons in the mean-field approximation to the chiral Lagrangian.

We have studied in particular the effects of medium modifications
of the vector meson properties on the dilepton invariant mass spectra.
Dilepton production has been calculated in two ways. In the form factor
approach, the formation time of the dileptons is neglected and the effects
of the intermediate vector mesons are included through
the electromagnetic form factors. In the dynamical approach, the
formation, propagation and decay of the intermediate vector mesons are
explicitly treated, and the change of the properties of these vector
mesons in the medium is included. It is
found that for vector mesons with large decay widths (e.g., rho meson), the
form
factor approach gives results  qualitatively similar to the
dynamical approach. However, for mesons with small decay widths
(e.g. phi meson), an explicit treatment of their formation, propagation
and decay is necessary, otherwise the medium effects will be
overestimated.

Differences are seen between dilepton spectra with and
without medium effects. These include the shift of the vector meson
peaks to lower
invariant masses  and the broadening of their widths.
For pion-pion annihilation,
the peak shifts to around 550 MeV, and its height increases by about a factor
of four. This peak is expected to be above the background from Dalitz
decays and bremsstrahlung which can in principle be subtracted from
the data. We thus conclude that
the in-medium properties of the rho meson can be studied through the
dilepton spectra. For kaon-antikaon annihilation, the situation
is more subtle. The peak shifts only slightly to a lower invariant
mass, so the experimental detection of this shift may be difficult.
In addition,  a shoulder appears around 950 MeV, which is
unfortunately only slightly
above the background from pion-pion annihilation.

As is well known, the  kaon-to-pion ratio increases with
increasing incident energies (at SIS energies, the $K^+/\pi ^+$ ratio is
about 1\%, while at AGS energies, this ratio increases to about 20\%).
It is thus expected that at higher incident energies (e.g., at AGS
energies), the shoulder in the kaon-antikaon contribution will be
well above the pion-pion contribution and therefore be easily detected
and more accurately analysed. To study this theoretically, we need,
however, first
to extend the relativistic transport model to include other
degrees of freedom so that it will be
suitable for the description of heavy-ion collisions at these energies.

\medskip
\noindent{\bf Acknowledgement}
\medskip
We are grateful to  M. Asakawa and Gy. Wolf for useful discussions.
This work was supported in part by the National Science Foundation
under Grant No. PHY-9212209
and the Welch Foundation under Grant No. A-1110.

\pagebreak
{\bf Figure Caption}

\vskip 0.5cm
{\bf Fig. 1:} ~~Time evolution of the average central density $\rho _/\rho _0$
(dotted curve), the pion, kaon, and
antikaon multiplicities. Dashed curves are the kaon and antikaon
multiplicities without medium effects.

{\bf Fig. 2:} ~~Effective masses of kaon, antikaon, rho meson and phi meson
as a function of density.

{\bf Fig. 3:} ~~Feynman diagram for dilepton production from  pion-pion
annihilation.

{\bf Fig. 4:} ~~Effective widths of  rho and phi meson as a function
of density.

{\bf Fig. 5:} ~~Feynman diagram for dilepton production from kaon-antikaon
annihilation.

{\bf Fig. 6:} ~~Dilepton invariant-mass spectrum from pion-pion
and kaon-antikaon annihilation without medium effects
in the dynamical approach.

{\bf Fig. 7:} ~~Same as Fig. 6 with medium effects.

{\bf Fig. 8:} ~~Comparison of dilepton spectra from  pion-pion annihilation
with and without medium effects.

{\bf Fig. 9:} ~~Same as Fig. 8 for kaon-antikaon annihilation.

{\bf Fig. 10:} ~~Same as Fig. 6 in the form factor approach.

{\bf Fig. 11:} ~~Same as Fig. 7 in the form factor approach.

\end{document}